\documentclass{emulateapj}

\shortauthors{Adelberger et al.}

\shorttitle{Optical Color-Selection of Galaxies at $1<z<3$}

\slugcomment{Received 2003 November 6; accepted 2004 January 21}

\newcommand{\secpoint}{\mbox{$''\mskip-7.6mu.\,$}}

\newcommand{\et}{{\it et al.}~}

\def\ltsima{$\; \buildrel < \over \sim \;$}
\def\simlt{\lower.5ex\hbox{\ltsima}}
\def\gtsima{$\; \buildrel > \over \sim \;$}
\def\simgt{\lower.5ex\hbox{\gtsima}}
\def\propsima{$\; \buildrel \propto \over \sim \;$}
\def\simprop{\lower.5ex\hbox{\propsima}}

\begin{document}
\title{OPTICAL SELECTION OF GALAXIES AT REDSHIFTS $1<Z<3$\altaffilmark{1}}

\author{\sc Kurt L. Adelberger\altaffilmark{2,3}}
\affil{Center for Astrophysics, 60 Garden Street, Cambridge, MA 02138}

\author{\sc Charles C. Steidel, Alice E. Shapley\altaffilmark{4},
	Matthew P. Hunt, Dawn K. Erb and Naveen A. Reddy}
\affil{Palomar Observatory, Caltech 105--24, Pasadena, CA 91125}

\author{\sc Max Pettini}
\affil{Institute of Astronomy, Madingley Road, Cambridge CB3 0HA, UK}

\altaffiltext{1}{Based in part on observations obtained at the W.M. Keck
Observatory, which is operated jointly by the California Institute of
Technology, the University of California, and NASA, and was
made possible by a gift from the W.M. Keck Foundation.}
\altaffiltext{2}{Harvard Society Junior Fellow}
\altaffiltext{3}{Current address: Carnegie Observatories, 813 Santa Barbara St, Pasadena, CA, 91101}
\altaffiltext{4}{Current address: Department of Astronomy, University of California, Berkeley, CA 94720}

\begin{abstract}
Few galaxies have been found between the redshift ranges $z\simlt 1$
probed by magnitude-limited surveys and $z\simgt 3$ probed by Lyman-break
surveys.  Comparison of galaxy samples at lower and higher redshift
suggests that large numbers of stars were born and the Hubble sequence
began to take shape at the intermediate redshifts $1<z<3$, but 
observational challenges have prevented us from observing the process
in much detail.  We present simple and efficient strategies that can
be used to find large numbers of galaxies throughout this
important but unexplored redshift range.  All the strategies
are based on selecting galaxies for spectroscopy on the basis
of their colors in ground-based images
taken through a small number of optical filters:
$G{\cal R}i$ for redshifts $0.85<z<1.15$,
$G{\cal R}z$ for $1<z<1.5$, and $U_nG{\cal R}$
for $1.4<z<2.1$ and $1.9<z<2.7$.  The performance of our strategies
is quantified empirically through spectroscopy of more than
2000 galaxies at $1<z<3.5$.  We estimate that more than half of the
UV-luminosity density at $1<z<3$ is produced by galaxies that satisfy
our color-selection criteria.  Our methodology is described in detail,
allowing readers to devise analogous selection criteria for
other optical filter systems.
\end{abstract}

\section{INTRODUCTION}
\label{sec:intro}

As the photons from the microwave background stream towards
earth they are gradually joined by other photons, first by those
produced in the occasional recombinations of the 
intergalactic medium, later by those cast off from cooling $H_2$ molecules,
later still, after hundreds of millions of years, by increasing
numbers
from quasars and galaxies, by bremsstrahlung
from the hot gas in galaxy groups and clusters,
and finally, shortly before impact,
by photons emitted
by the Milky Way's own gas and stars and dust.  All reach
the earth together.  They
provide a record of the
history of the universe throughout its evolution, but it is
a confused record. Two photons
that simultaneously pierce the same pixel of a detector may have
been emitted billions of years apart by regions of the
universe that were in vastly different stages of evolution.
One of the challenges in observational
cosmology is to separate the layers of history that
we on earth receive superposed.

This paper is concerned with a small part of the challenge:
locating galaxies at redshifts $1\simlt z\simlt 3$ among
the countless 
objects that speckle the night sky.
The redshift range
is interesting for a number of reasons.  During the
$\sim 4$ billion years that elapsed between redshifts~3
and~1, the comoving density of star formation was
near its peak, the Hubble sequence of galaxies may have been established,
and a large fraction of the stars in the universe were born (Dickinson et al. 2003, Rudnick et al. 2003).
Observing star-formation at these redshifts will likely play an important
role in our attempts to understand galaxy formation.

Identifying galaxies at $1<z<3$ would be easy if observing time were infinite.
One could simply obtain a spectrum of every object in a deep image and
discard those at lower or higher redshift.
In practice telescope time is limited
and one would like to spend as much of it as possible
studying
objects of interest---not finding them.  
Our aim is to present
simple strategies that allow observers to locate star-forming galaxies
at $1<z<3$ with only a
minimal loss of telescope time to
deep imaging or spectroscopy of objects at the wrong redshifts.

We shall assume throughout that the goal is to identify star-forming
galaxies that lie inside a narrow range of redshifts within the wider
span $1<z<3$.  Concentrating on a narrow range
has obvious observational advantages;
galaxies at similar redshifts
have similar spectral features that lie at similar observed wavelengths,
and consequently the spectrograph can be chosen and configured in a way that is
optimized for every object on a multislit mask.
It is also necessary for many scientific programs.
One might want to discover where galaxies
lie relative to the intergalactic gas that produces absorption
lines in the spectra of a QSO at $z\sim 2.5$, for example, or
to create a large and homogeneous sample
of galaxies at a single epoch in the past that can then be
compared to existing samples of galaxies in the local universe.
Projects with wildly different goals may benefit less from
the strategies we advocate.  Standard magnitude-limited spectroscopy or
photometric redshifts might be preferable if (for example) 
one wanted to create a sample that contained all galaxy
types at all redshifts.

Our approach exploits
the
fact that 
the wavelength-dependent cross-sections
of various common atoms and ions
give distinctive
colors to galaxies at
different redshifts. 
It has been recognized for decades that measuring a galaxy's brightness
through a range of broad-band filters should therefore provide some
indication of its redshift
(e.g., Baum 1962; Koo 1985; Loh \& Spillar 1986).
Recent work suggests that a redshift accuracy of $\sigma_z\simlt 0.1$
can be achieved for galaxies at $0\simlt z\simlt 6$ given
extravagantly precise photometry
through 7 filters that span
the wavelength range
$0.3\simlt\lambda\simlt 2.2\mu$m (e.g., Hogg et al. 1998; 
Budav\'ari et al. 2000; Fern\'andez-Soto et al. 2001; Rowan-Robinson 2003).
An accuracy $\sigma_z\simlt 0.1$ 
is clearly sufficient for finding
galaxies at $1<z<3$,
but obtaining the necessary deep images through numerous filters
consumes enormous amounts of
telescope time that could be more profitably devoted to galaxy spectroscopy.
We were led to
seek specialized color-selection techniques that could identify galaxies
at $1<z<3$ even in comparatively noisy images taken through a small number
of optical filters.
\footnote{As readers will notice, the samples created by our techniques
will hardly differ from those that would be created if standard
photometric-redshift techniques were applied to the same
noisy two-color data; our
goal is not to disparage photometric-redshift 
techniques but to adapt them to a new regime.}

\begin{figure}[htb]
\centerline{\epsfxsize=9cm\epsffile{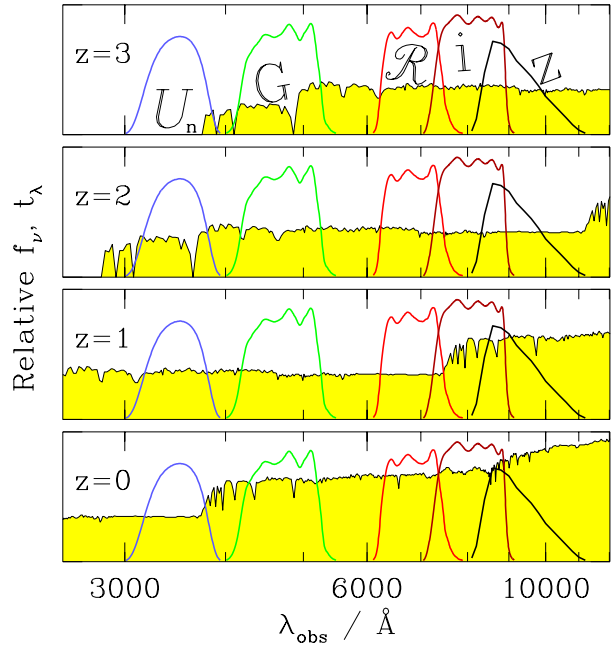}}
\figcaption[f1.eps]{ $U_nG{\cal R}iz$ colors of galaxies at redshifts $0<z<3$.
Shaded curves show the spectrum of a model star-forming galaxy
(type Im from \S~\ref{sec:bcseds}) at various redshifts.
Unshaded curves show the transmission curves of the 
filters used in this paper.
\label{fig:filters}
}
\end{figure}

Such techniques can succeed only if they take advantage of
strong and obvious features in galaxies' spectra.
No feature is stronger
than the Lyman break at 912\AA\ produced by
the photo-electric opacity of hydrogen in its ground state.
Meier (1976) argued that the strength of this break would
allow high-redshift galaxies to be identified in images
taken through just three filters, a claim that Steidel et al. (1996; 1999)
have since confirmed.  Although the Lyman break itself is not visible
from the ground at $z\simlt 3$, other weaker features are,
and the success of two-color selection at $z>3$ 
inspired us try to develop similar two-color optical selection
strategies for $1<z<3$.
Figure~\ref{fig:filters} shows 
some of the spectral features that
we had to work with.  Most obvious, after the Lyman-break,
is the Balmer-break at 3700\AA.  The strength of the Balmer-break
can be estimated from the model galaxy spectra described in \S~\ref{sec:bcseds}
or
from the galaxy observations described in \S~\ref{sec:obs}.  
\S~\ref{sec:bbg} explains
how it can be used to find galaxies at $1\simlt z\simlt 1.5$.
Franx et al. (2003) and Davis et al. (2003) have also used this feature to
find distant galaxies.
At $1.5\simlt z\simlt 2.5$ no strong breaks are present in the optical
spectra of galaxies, but the lack of spectral breaks is itself
a distinguishing characteristic of
galaxies at these
redshifts. \S\S~\ref{sec:bx},~\ref{sec:bmz}, and~\ref{sec:anyz} explain.
Our results are summarized and discussed in \S~\ref{sec:summary}.
Together with the Lyman-break technique,
the selection techniques presented here
allow the efficient creation
of large samples of star-forming galaxies throughout
the redshift range $1\simlt z\simlt 5$. 

\section{MODEL GALAXY SPECTRA}
\label{sec:bcseds}
Our development of selection strategies began with theoretical models of
galaxy spectra.  At redshifts $z\simlt 3$ a galaxy's optical broad-band colors
are determined by the mixture of stellar types it contains.  This is in turn
largely determined by a galaxy's star-formation history. Galaxies that formed
most of their stars recently have spectra dominated by bright and hot
massive stars, while galaxies that formed most of their stars in the distant
past will have spectra dominated by fainter, cooler, but longer-lived low-mass stars.
We considered five model galaxy spectra that were intended to span the range
of possible star-formation histories.  Each model spectrum was calculated with
the code of Bruzual \& Charlot (1996, private communication) and assumed that the galaxy's star-formation
rate as a function of time, $S(t)$, was a decaying exponential: $S(t)\propto \exp(-t/\tau)$.
The adopted values of $\tau$ and assumed time-lapse since the onset of star formation
for our five models are listed in Table~\ref{tab:bcseds}.  Bruzual \& Charlot (1993)
show that star-formation histories with these parameters reproduce the observed
spectra of different galaxy types in the local universe.  Because a wide range of
star-formation histories can result in nearly identical model spectra, we were
not concerned that some of our adopted model parameters are physically impossible
at high redshift due to the young age of the universe.  
Parameter combinations that are more plausible can produce similar spectra,
and in any case our aim was only to have model spectra that roughly 
spanned the range of conceivability.  Subsequent empirical refinements of our selection
criteria would compensate for any shortcomings in our model spectra.

\begin{deluxetable}{lrr
}\tablewidth{0pc}
\scriptsize
\tablecaption{Model SED Parameters}
\tablehead{
	\colhead{Name} &
        \colhead{$\tau$\tablenotemark{a}} &
        \colhead{age\tablenotemark{a}} 
}
\startdata
E	&	1	& 13.8 	\\
Sb	&	2	&  8.0	\\
Sbc	&	4	& 10.5	\\
Sc	&	7	& 12.3	\\
Im	&  $\infty$	&  1.0  \\
\enddata
\tablenotetext{a}{Gyr}
\label{tab:bcseds}
\end{deluxetable}

We estimated the colors of galaxies
at different redshifts by scaling the wavelengths of our template spectra by $1+z$, applying the appropriate
amount of absorption due to intergalactic hydrogen (Madau 1995), and finally
multiplying the result by our filter transmissivities.  In some cases, mentioned explicitly below, the template
galaxies were first reddened by dust that followed a Calzetti (1997) attenuation law.  
For the reddening $E(B-V)=0.15$ that appears typical for high-redshift
galaxies (Adelberger \& Steidel 2000) the resulting change in galaxy
color is not large:  0.22 magnitudes from rest-frame 1500\AA\ to 2000\AA,
0.16 from 2000\AA\ to 2500\AA, 0.13 from 2500\AA\ to 3000\AA,
0.21 from 3000\AA\ to 4000\AA.
The formulae that we used to calculate template galaxies' colors can be found (e.g.)
in \S~4.2 of Papovich, Dickinson, \& Ferguson 2001.

\section{OBSERVATIONS}
\label{sec:obs}
Although our initial ideas for
color-selection strategies were motivated by
model galaxy spectra (\S~\ref{sec:bcseds}),
our final selection
criteria were determined by the observed broadband colors of
galaxies at different redshifts.

The galaxies we used lay within fields
observed during our survey of Lyman-break galaxies
at $z\sim 3$ (Steidel et al. 2003) or within fields
chosen with similar criteria.
$U_nG{\cal R}$ images of each field were
obtained as described in Steidel et al. (2003).
In some cases these images
were supplemented with $i$ band images to increase our wavelength
coverage.
In one field, the HDF-North, a publicly released $z$ image from the
GOODS survey (Dickinson \& Giavalisco 2002) gave us photometric coverage of the entire optical
window.  Transmission curves for these filters are shown
in Figure~\ref{fig:filters}.  
The typical set of $U_nG{\cal R}i$ images was
taken in $\sim 1''$ seeing at the Palomar $200$-inch telescope with
exposure times (median) of $7.0$, $2.3$, $1.8$, and $2.0$ hours
respectively.  All reported magnitudes are in the AB system.
A complete description of our $U_nG{\cal R}i$ observations 
can be found in Steidel et al. (2003).  

Galaxies were selected for spectroscopic follow-up on the basis
of their colors.  Initially our spectroscopy
was exploratory, as we sought to establish whether a certain
combination of colors reliably indicated that
a galaxy lay within a targeted redshift range.  During this
phase galaxies were observed on multislit masks that were
primarily devoted to the Lyman-break survey.  The spectroscopic
set-up for these observations ($1\secpoint 4$ slits, $\sim 10$\AA\ resolution,
$4000\simlt\lambda\simlt 7000$\AA\ wavelength coverage,
$\sim 1.5$ hour integration time with LRIS on the Keck I or II telescopes) 
is described in Steidel et al. (2003).
Later, after 
our initial color-selection criteria had been
validated or refined, entire masks were filled with objects that satisfied them,
and the spectroscopic set-up was optimized
to the targeted redshift.  For redshifts $1.5<z<2.5$, where our
redshift measurements were based primarily on absorption lines
in the rest-frame far-UV, the difference from the Lyman-break survey
set-up was slight.  Most of these spectra were obtained with 
the blue arm of LRIS using a 400 line mm$^{-1}$ grism
blazed at 3400\AA\ and
$1\secpoint 2$ slits.   The resulting spectra
had $\sim 5$\AA\ resolution
and stretched from $3100$ to $8000$\AA.
For roughly half of the spectra we concurrently obtained spectra
with the red arm, usually with a 6800\AA\ dichroic
and 400 line mm$^{-1}$ $8500$\AA\ grating that provided spectra of $\sim 5$\AA\
resolution from 6800\AA\ to 9500\AA.
At $z\sim 1$ our redshift
measurements were based primarily on the [OII]$\lambda 3727$ doublet.
Here we benefited from redder, higher-resolution gratings
and could tolerate shorter exposure times.  $0\secpoint 8$ slits,
$\sim 3$\AA\ resolution, wavelength coverage from $6300$\AA\ to
$8700$\AA, and 1 hour integration times were typical.
All data were reduced with procedures similar to those
described in Steidel et al. (2003).

\begin{figure}[htb]
\centerline{\epsfxsize=9cm\epsffile{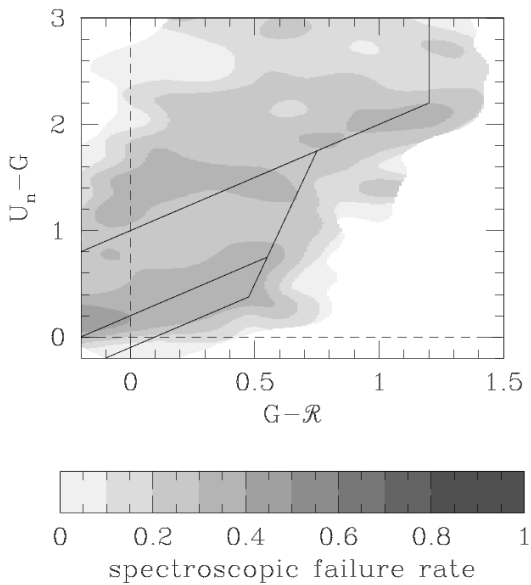}}
\figcaption[f2.eps]{The fraction of spectroscopically observed galaxies for which we were unable to measure
a redshift as a function of $U_nG{\cal R}$ color.
\label{fig:specsuccess}
}
\end{figure}

Because our goal was to measure the maximum number of galaxy redshifts,
we would begin observing a new slitmask after the previous slitmask's allotted
integration time had elapsed, even if (as was usually the case)
our exposures had not been long enough to produce spectra that allowed
us to measure a redshift for every object.
In the best conditions we were able to
measure redshifts for $\simgt 90$\% of
the objects.  In the worst conditions few of our spectra were usable.
Averaging over all conditions, our net spectroscopic success rate
was roughly $75$\%.  We have been unable to find any evidence 
that objects with identified and unidentified spectra lie at significantly different
redshifts.  In the numerous cases where we were able to measure
a previously unidentified object's redshift by observing it again on
a new slitmask, its redshift was similar to those of the objects
whose spectra were identifiable after the first attempt.
Nevertheless readers should be aware that our sample suffers from
some incompleteness.  Figure~\ref{fig:specsuccess} shows our spectroscopic
failure rate as a function of $U_nG{\cal R}$ color for objects
whose spectra were obtained with the blue spectroscopic configuration
described above and in Steidel et al. (2003).  There is little evidence
that the failure rate depends strongly on objects'
intrinsic colors within the color selection windows described below.

\section{COLOR SELECTION AT $1.0\simlt Z\simlt 1.5$}
\label{sec:bbg}
\subsection{Balmer-break selection at $z\simeq 1.0$}
\begin{figure}[htb]
\centerline{\epsfxsize=9cm\epsffile{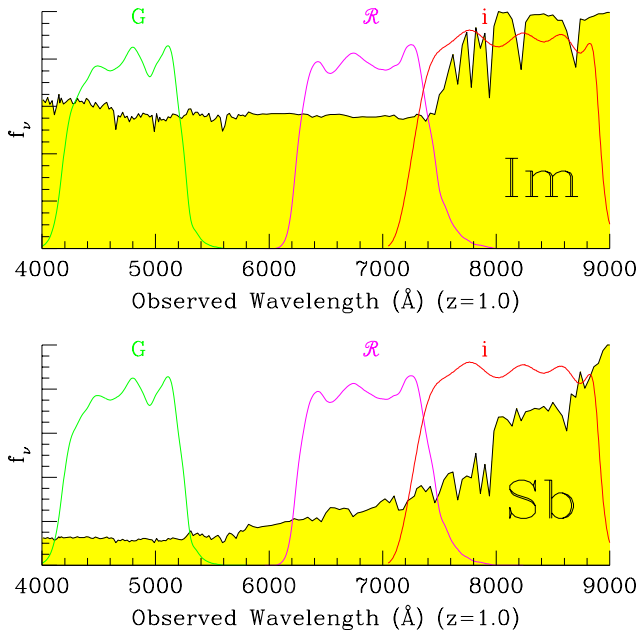}}
\figcaption[f3.eps]{
The Balmer and 4000\AA\ breaks.  Younger galaxies, represented by the
``Im'' model spectrum (\S~\ref{sec:bcseds}), have a Balmer break near
3700\AA\ due to Hydrogen Balmer absorption in the spectra of
B, A, and F stars.  Older galaxies, represented by the ``Sb'' model
spectrum (\S~\ref{sec:bcseds}), have 4000\AA\ CaII H\&K breaks.  
These breaks give distinctive $G{\cal R}i$ colors to galaxies at $z\sim 1$.
\label{fig:bbgmethod}
}
\end{figure}

Most galaxies have a
break in their spectra at $\lambda_{\rm rest}\sim 4000$\AA\
that is produced by a combination of Hydrogen Balmer-continuum absorption
in the spectra of B, A, and F stars and CaII H\&K absorption
in the spectra of F, G, and K stars.  The relative strengths
of the Balmer and 4000\AA\ breaks depends upon the mixture of
stellar types in a galaxy---younger galaxies have stronger
Balmer breaks and older galaxies have stronger 4000\AA\ breaks---but
few galaxies have no break at all
(Figure~\ref{fig:bbgmethod}).

\begin{figure}[htb]
\centerline{\epsfxsize=9cm\epsffile{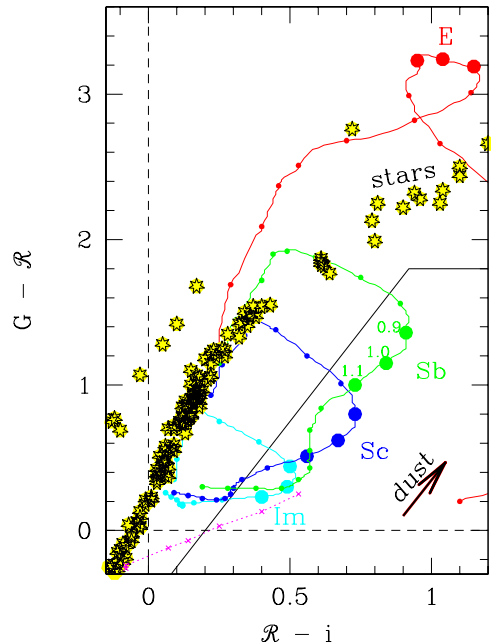}}
\figcaption[f4.eps]{
Expected locations of stars and galaxies on a $G{\cal R}i$ two-color diagram.
The stars are from Gunn \& Stryker (1983).  The curved tracks show the colors
of model galaxies of different spectral types (see \S~\ref{sec:bcseds}) at
various redshifts, starting at $z=0$ and increasing clockwise.  Circles
mark redshift intervals of $dz=0.1$; large circles mark $z=0.9, 1.0, 1.1$.
Dust reddening (Calzetti 1997) to $E(B-V)=0.15$ was assumed.  
The arrow indicates the color change corresponding to $\Delta E(B-V)=0.2$
for an Sc
galaxy at $z=1.0$.
One would expect to find galaxies at $z\sim 1$ and little else to the
right of the diagonal line.  Since the line is nearly parallel to the
reddening vector, this conclusion does not depend strongly on galaxies'
assumed dust content.  Sb galaxies fall to the right of the line even for redshifts
significantly larger than $z=1$, but higher-redshift Sbs are not a major
component of our sample because Sb spectral types are significantly
less common than Sc at $z\sim 1$ (cf. figure~\ref{fig:showcandsz1})
and because higher redshift sources are fainter and less
easily able to satisfy any apparent-magnitude limit.
The dotted line shows the colors of a model
galaxy at $z=1.0$ that has been forming stars at a constant rate. Crosses mark
its colors after 3,10,30,100,300,1000 Myr of star-formation.  The galaxy becomes
redder as it ages and begins to satisfy our selection criteria for $t\simgt 30$Myr.
\label{fig:expcolz1}
}
\end{figure}

Figure~\ref{fig:filters} suggests that this break should
give distinctive broadband colors to galaxies at $z\sim 1$.
At no other redshift will a strong break between ${\cal R}$
and $i$ be accompanied by a flatter spectrum at bluer wavelengths.
Our first guess at
photometric selection criteria targeting 
galaxies at $z\sim 1$
was inspired by figure~\ref{fig:expcolz1},
which shows the $G{\cal R}i$ colors of model galaxies at
various redshifts (\S\ref{sec:bcseds}) and of stars in our own galaxy.
Spectroscopic
follow-up of
objects with colors that are characteristic of $z\sim 1$ galaxies and
distinct from the colors of other objects 
should produce
a redshift survey consisting primarily of galaxies at $z\sim 1$.
These objects lie to the right of the diagonal line
in figure~\ref{fig:expcolz1}, suggesting
\begin{equation}
{\cal R}-i \geq 0.4(G-{\cal R}) + 0.2,\quad\quad G-{\cal R}\leq 1.8 \quad\quad\quad\mathbf{[FN]}
\label{eq:fn}
\end{equation}
as reasonable criteria for identifying the $z\sim 1$ galaxies
in deep $G{\cal R}i$ images.   These criteria are called ``FN''
in our internal naming convention and may be referred to by that name
in subsequent publications.

Even if galaxies at $z\sim 1$ had colors that matched the models
perfectly, and even if we suffered no photometric errors, figure~\ref{fig:expcolz1}
makes it clear that some galaxies at $z\sim 1$ would not satisfy
the selection criteria above.  These are galaxies
with extreme star-formation histories.  At one extreme are
galaxies whose present star-formation rates are much lower
than their past average.  
A survey of star-forming galaxies at $z\sim 1$
(such as ours) will be only negligibly affected by their omission.
Of more potential concern are galaxies at the opposite extreme, galaxies
with present star-formation rates much higher
than their past average.  The spectra of these galaxies are
dominated by light from massive O stars---by stars whose atmospheres are
too hot to contain significant amounts of neutral Hydrogen---and
consequently they do not have Balmer breaks.  Since a Balmer break
begins to be discernible when a star-formation episode has lasted
longer than the typical $\sim 10^7$ yr lifetime of an O star,
and since most star formation in the local universe (e.g., Heckman 1997)
and at $z\sim 3$ (e.g., Papovich et al. 2001; Shapley et al. 2001)
occurs in episodes that last substantially longer than $10^7$ yr,
we suspected that our reliance on the Balmer-break in our selection criteria
would not cause us to miss significant numbers of star-forming galaxies
at $z\sim 1$.

\begin{figure}[htb]
\centerline{\epsfxsize=9cm\epsffile{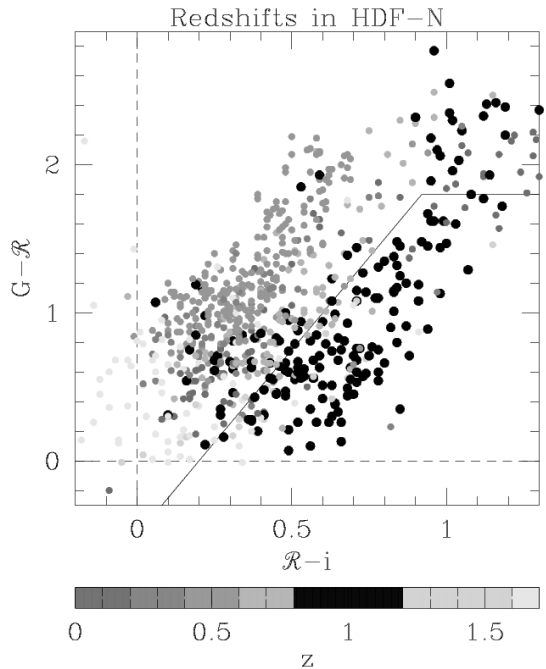}}
\figcaption[f5.eps]{
Observed redshifts of objects with different $G{\cal R}i$ colors.
The data are taken from various magnitude-limited surveys in the
HDF-N; see text.  
Known galaxies at $z\sim 1$ have colors similar to our expectations from
figure~\ref{fig:expcolz1}.
The solid line shows the $z\sim 1$ selection criteria of equation~\ref{eq:fn}.
\label{fig:showcandsz1}
}
\end{figure}

In order to obtain empirical support for our proposed selection
criteria (equation~\ref{eq:fn}), we began in August 1995
to obtain
$G{\cal R}i$ images in fields with completed or
ongoing magnitude-limited redshift surveys.
Four fields were chosen: the 00$^h$53 field of Cohen \et (1999),
the Hubble Deep Field (Williams \et 1996; Cohen \et 2000), the 14$^h$18 field
of Lilly \et (1995), and the 22$^h$18 field of Cowie \et (1996)
and Lilly \et (1995).   Spectroscopic redshifts have been
published for 1312 objects in these fields combined.  The published
coordinates for these objects were sufficient for us to
easily and unambiguously identify 1138 of them with
objects in our images.  738 of the objects are in the HDF-N,
where spectroscopic follow-up has been especially deep and complete;
their
colors and redshifts are shown in
figure~\ref{fig:showcandsz1}.  One can see that the bulk of known
galaxies at $z\sim 1$ 
have $G{\cal R}i$ colors
that satisfy our selection criteria.

The galaxies at $0.85<z<1.15$ that do not satisfy our $G{\cal R}i$
selection criteria can be crudely grouped into three classes.
Please refer to figure~\ref{fig:showcandsz1}.  First,
there are the handful of galaxies with
${\cal R}-i\simeq 0.9$, $G-{\cal R}\simgt 2.0$.  These galaxies
have rest-frame colors nearly identical to those of
local ellipticals (see figure~\ref{fig:expcolz1}).  They
are objects that formed the bulk of their stars in the past
and are no longer forming stars at a significant rate.  Their
absence from a color-selected survey of star-forming
galaxies is expected and harmless.  Second, there are
objects with colors identical to those of low redshift galaxies
but with reported spectroscopic redshifts $z\simeq 1.0$.
Some of these objects may have unusual star-formation histories
or large photometric errors
or exceptionally strong emission lines,
and some may have incorrectly measured spectroscopic redshifts.
Third, there are objects with measured colors that lie
just outside the Balmer-break color selection window.
These galaxies may have been scattered out of our selection
window by photometric errors, which are typically $\sim 0.2$
magnitudes in both $G-{\cal R}$ and ${\cal R}-i$.  Their
absence from a Balmer-break selected survey can be
largely corrected with statistical techniques
described in Adelberger (2002,2004).

\begin{figure}[htb]
\centerline{\epsfxsize=9cm\epsffile{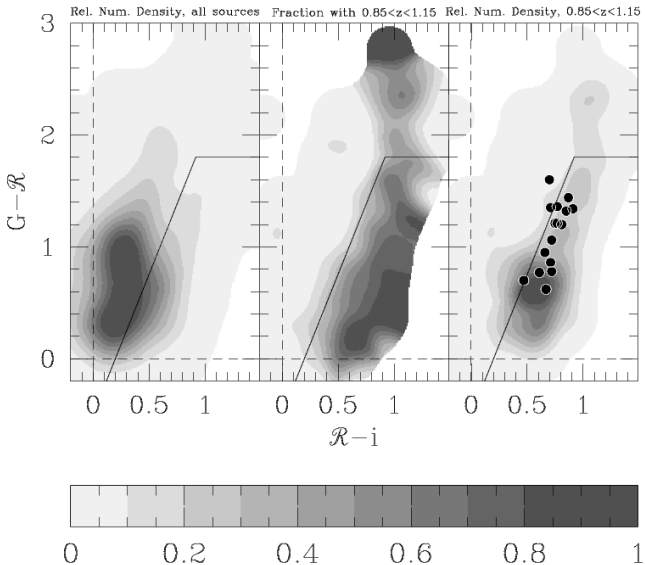}}
\figcaption[f6.eps]{Left panel:  The distribution of colors for all objects in deep $G{\cal R}i$ images
of the HDF-N.  Middle panel:  The probability that an object has $0.85<z<1.15$ as a function
of $G{\cal R}i$ color.  Right panel:  The distribution of $G{\cal R}i$ colors for
objects with $0.85<z<1.15$ in the HDF-N.  Superimposed are the colors of
the 16 ISO $15\mu$m sources with $0.84<z<1.15$ in this field (see Adelberger \& Steidel 2000 and references
therein; 41 of the 46 ISO sources in this field have measured spectroscopic
redshifts, so it is unlikely that there is a large number of
spectroscopically unidentified ISO sources at $z\sim 1$ with
colors significantly different from these).  These extremely dusty sources are among the most rapidly star-forming galaxies
at $z\sim 1$, but their optical colors are sufficiently similar to those of more normal galaxies
that they satisfy our selection criteria (equation~\ref{eq:fn}, solid line in all panels).  
There is little evidence that our UV-based selection
criteria bias our samples against the dustiest galaxies, a point that has been more laboriously
made by Adelberger \& Steidel (2000) and Adelberger (2001).
\label{fig:z1triptych}
}
\end{figure}

Figure~\ref{fig:z1triptych} presents the data of figure~\ref{fig:showcandsz1}
in a way that may be easier to grasp.  We produced one alternate
realization of figure~\ref{fig:showcandsz1}'s data by adding
to each galaxy's $G{\cal R}i$ colors a Gaussian deviate with standard
deviation $\sigma=0.1$ that is similar to the color's measurement
uncertainty.  After repeating this procedure numerous times,
we concatenated the alternate realizations into a large list
of $G-{\cal R}$, ${\cal R}-i$, $z$ triplets, and then calculated,
for each point in the $G{\cal R}i$ plane, the fraction of galaxies
with those colors in the large list that had redshift $0.85<z<1.15$.
Figure~\ref{fig:z1triptych} shows the result.  
As anticipated, the portion of the $G{\cal R}i$
plane that is dominated by star-forming galaxies at $z\sim 1$
is approximately described by equation~\ref{eq:fn}.  
One exception is the region near
$G-{\cal R}\sim 2.8$, ${\cal R}-i\sim 0.9$, which is populated primarily by
galaxies at $z\sim 1$ but lies outside our selection window.
It would be easy to modify the window to include this region,
but, as mentioned above, galaxies at $z\sim 1$ with these colors 
do not account for much of the star-formation density.  We chose to ignore them.
Readers interested in stellar mass rather than star-formation rate
might choose differently.

Ideal color-selection criteria would be perfectly complete and
perfectly efficient; they would be satisfied by every galaxy
in the targeted redshift interval and only by galaxies in the
targeted redshift interval.  Figure~\ref{fig:z1triptych} shows
that in practice the goals of completeness and efficiency are
incompatible.  Photometric errors and intrinsic variations in the spectra
of galaxies cause galaxies inside the redshift interval $0.85<z<1.15$
to have a wide range of $G{\cal R}i$ colors.  In some cases
these colors are identical to those of galaxies at other redshifts.
If we wanted our color-selected survey to be as complete as possible, 
we would want to make our selection box very large so that it would
include even galaxies with large photometric errors or abnormal spectral
shapes, but this improvement in completeness would come at the price
of admitting more galaxies at the wrong redshifts and it would
therefore decrease our efficiency.  To quantify how closely
our selection criteria satisfied the conflicting goals
of completeness and efficiency, we used the magnitude-limited surveys
discussed above to calculate two quantities.
$\alpha$, the completeness coefficient,
is equal to the fraction of galaxies at $0.85<z<1.15$ in the magnitude limited
surveys whose colors satisfied equation~\ref{eq:fn}.
$\beta$, the efficiency coefficient, is equal to the fraction of
magnitude-limited survey objects in the selection box of equation~\ref{eq:fn}
whose measured redshift satisfied $0.85<z<1.15$.

Table~\ref{tab:fncompint} lists $\alpha$ and $\beta$, with and without
weighting by the galaxies' apparent luminosities through various filters.
Our completeness depends 
on wavelength.
Samples selected through equation~\ref{eq:fn} are
especially complete for the bluest galaxies:
the $U_n$-weighted column shows that approximately $85$\% of the $U_n$
(rest-frame $1800$\AA) luminosity density detected
at $0.85<z<1.15$ in magnitude-limited surveys
is produced by galaxies that satisfy
the selection criteria.  The completeness falls
towards redder wavelengths, where the total $z\sim 1$ luminosity
density receives larger contributions from older galaxies
whose spectra are less dominated by star formation,
but even at rest-frame $4500$\AA\ (observed $z$-band)
approximately $70$\% of the luminosity density at $0.85<z<1.15$
is produced by galaxies whose colors satisfy equation~\ref{eq:fn}.
Because our color-selected catalogs extend to magnitudes significantly
fainter than those of the magnitude-limited surveys,
the detected luminosity density at $z\sim 1$ in our survey
is far higher than in the magnitude-limited surveys.  The completeness
fractions above apply only to the magnitude range ${\cal R}\simlt 24$
where the surveys overlap. 

\begin{deluxetable*}{lcccclcc
}\tablewidth{0pc}
\scriptsize
\tablecaption{Color-selection Efficiency, $1\simlt z\simlt 1.5$}
\tablehead{
	\colhead{Name} &
        \colhead{$\alpha,\beta$\tablenotemark{a}} &
        \colhead{$\alpha_U,\beta_U$\tablenotemark{b}} &
        \colhead{$\alpha_i,\beta_i$\tablenotemark{c}} &
        \colhead{$\alpha_z,\beta_z$\tablenotemark{d}} &
        \colhead{Reference} &
	\colhead{$N_{\rm cand}$\tablenotemark{e}} &
	\colhead{$N_{\rm z}$\tablenotemark{f}}
}
\startdata
$0.85<z<1.15$:&               &               &               &           &               &                  \\
00$^h$53      & 0.73,0.55 & 0.95,0.55 & 0.77,0.55 &           & Cohen \et 1999 & 34 & 168\\
HDF-N         & 0.67,0.55 & 0.81,0.57 & 0.67,0.49 & 0.65,0.51 & Cohen \et 2000 & 166 & 676\\
14$^h$18      & 0.53,0.67 & 0.60,0.56 & 0.57,0.64 &           & Lilly \et 1995 & 15 & 123\\
22$^h$18      & 0.65,0.46 & 0.87,0.67 & 0.67,0.50 &           & Cowie \et 1996 & 24 & 171\\
median        & 0.66,0.55 & 0.84,0.57 & 0.67,0.53 & 0.65,0.51 &                &    &    \\
$1.0<z<1.5$:  &               &               &               &           &               &                  \\
HDF           & 0.80,0.53 & 0.84,0.60 & 0.78,0.53 & 0.77,0.54 & Cohen \et 2000 & 100 & 676\\
\enddata
\tablenotetext{a}{The fraction of galaxies with ${\cal R}>21$ at the redshift of interest whose colors satisfy our proposed selection criteria ($\alpha$) and the fraction of objects with ${\cal R}>21$ satisfying our selection criteria whose redshift lies in the desired range ($\beta$).}
\tablenotetext{b}{$\alpha$ and $\beta$ recalculated after assigning each galaxy a weight proportional to its apparent $U_n$ luminosity.}
\tablenotetext{c}{$\alpha$ and $\beta$ recalculated after assigning each galaxy a weight proportional to its apparent $i$ luminosity.}
\tablenotetext{d}{$\alpha$ and $\beta$ recalculated after assigning each galaxy a weight proportional to its apparent $z$ luminosity.}
\tablenotetext{e}{Number of spectroscopic redshifts in redshift range of interest (${\cal R}>21$)}
\tablenotetext{f}{Total number of spectroscopic redshifts (${\cal R}>21$)}
\label{tab:fncompint}
\end{deluxetable*}

\begin{figure}[htb]
\centerline{\epsfxsize=9cm\epsffile{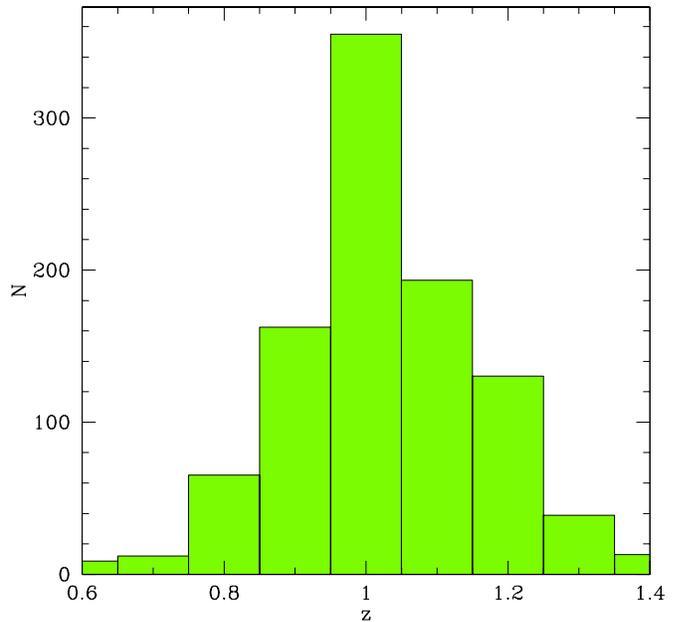}}
\figcaption[f7.eps]{
The redshift histogram the spectroscopically observed sources  in our fields
whose colors satisfied the selection criteria of equation~\ref{eq:fn}.
\label{fig:bbgnz}
}
\end{figure}

In the fall of 1997
we began to add occasional $z\sim 1$ galaxy candidates
to our Lyman-break galaxy slitmasks.  By the spring of 1999 we
had settled on our selection criteria (equation~\ref{eq:fn}) and
were devoting entire slitmasks to Balmer-break galaxies.
Figure~\ref{fig:bbgnz}
shows the overall redshift distribution of the objects
observed to date.  Excluding the handful of galaxies with $z<0.4$,
whose unexpected colors usually resulted from abnormally strong
nebular emission lines, the mean redshift is $\langle z\rangle=1.02$
and the r.m.s. is $\sigma_z=0.15$.

\subsection{Balmer-break selection at higher redshifts}
In principle it would
be easy to use similar color-selection criteria
to find galaxies at almost arbitrarily high redshifts.
In practice there is little reason to pursue this selection strategy
beyond the redshift $z\sim 1.5$ where the Balmer-break
leaves the optical window; near this redshift bluer spectral features
are beginning to enter the optical window and by exploiting these 
one can continue to take advantage of well-developed CCD detector
technology.

\begin{figure}[htb]
\centerline{\epsfxsize=9cm\epsffile{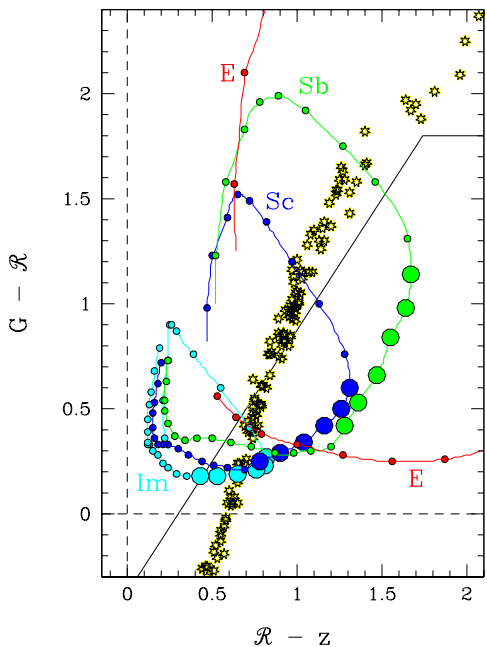}}
\figcaption[f8.eps]{
The expected $G{\cal R}z$ colors of stars and galaxies.  The symbols are
as in figure~\ref{fig:expcolz1}, except here large dots mark the colors
of model galaxies with $E(B-V)=0.15$ 
at $z=1.0$, 1.1, 1.2, 1.3, 1.4, 1.5 and the solid line
shows the selection criteria of equation~\ref{eq:fnz15}.
\label{fig:expcolsgrz}
}
\end{figure}

\begin{figure}[htb]
\centerline{\epsfxsize=9cm\epsffile{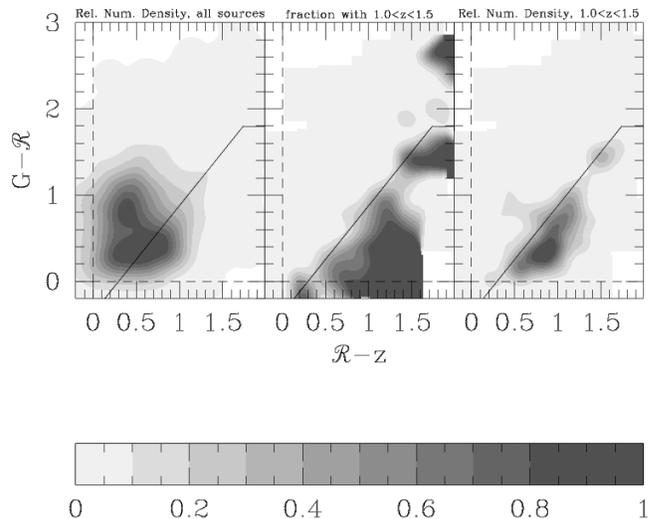}}
\figcaption[f9.eps]{
Similar to figure~\ref{fig:z1triptych}, except results are shown
for $G{\cal R}z$ selection of galaxies with $1.0<z<1.5$ rather
than $G{\cal R}i$ selection of galaxies with $0.85<z<1.15$.
\label{fig:z15triptych}
}
\end{figure}

Little thought is required to extend the Balmer-break selection to $z\sim 1.5$.
Consider figure~\ref{fig:expcolsgrz}, for example, which shows
the expected $G{\cal R}z$ colors of galaxies at different redshifts.
The color cuts
\begin{equation}
{\cal R}-z \geq 0.8(G-{\cal R}) + 0.3 \quad\quad\quad G-{\cal R}<1.8
\label{eq:fnz15}
\end{equation}
isolate galaxies at $1.0<z<1.5$ from the foreground and background
populations.  Figure~\ref{fig:z15triptych}, calculated in an analogous manner
to figure~\ref{fig:z1triptych}, confirms that objects in the HDF with these colors
tend to lie at redshifts $1.0<z<1.5$.  Table~\ref{tab:fncompint}
lists the completeness parameters $\alpha$ and $\beta$ for
these selection criteria.  Approximately
$80$\% of the known luminosity density at redshifts $1.0<z<1.5$
and observed wavelengths $3000\simlt\lambda\simlt9500$\AA\ in the HDF
is produced by galaxies whose colors satisfy equation~\ref{eq:fnz15}.
These numbers should be treated with some caution, since the
incompleteness of the HDF magnitude-limited spectroscopy strongly skews the
observed redshift distribution towards the lower
end of the range $1.0<z<1.5$.  Nevertheless we hope to have demonstrated that
Balmer-break selection allows one to create reasonably complete catalogs
of star-forming galaxies throughout the redshift range $1.0\simlt z\simlt 1.5$.


\section{COLOR SELECTION AT $1.9\simlt Z\simlt 2.7$}
\label{sec:bx}
Figure~\ref{fig:filters} shows that the absence of a strong break in the
optical window is a distinguishing characteristic of galaxies
at $z\sim 2$.  Ruling out the existence of a break requires photometry
through at least the five filters shown in the figure, but we wanted
to devise selection criteria that required imaging through only
three.  We chose to use the $U_nG{\cal R}$ filters
for reasons of convenience; our ongoing Lyman-break
galaxy survey was producing numerous deep $U_nG{\cal R}$ images
that we wanted to use for other purposes.

\begin{figure}[htb]
\centerline{\epsfxsize=9cm\epsffile{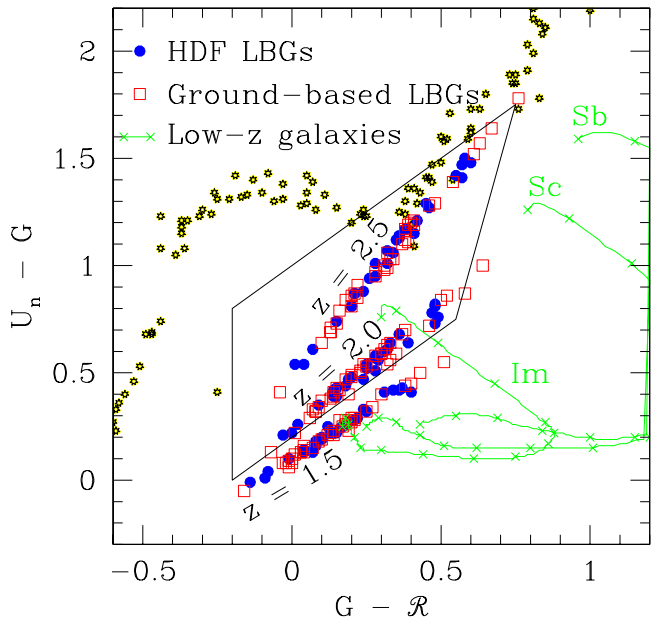}}
\figcaption[f10.eps]{
The expected $U_nG{\cal R}$ colors of stars and of star-forming galaxies at 
various redshifts.  Circles and squares
mark the colors that would result from shifting the observed SEDs of
Lyman-break galaxies at $z\sim 3$ to $z=1.5$, 2.0, 2.5.  Data are from
Papovich et al. (2001; HDF) and Shapley et al. (2001; ground-based).  Curved tracks show
the expected colors of model galaxies (\S~\ref{sec:bcseds}) at lower redshifts, starting at
$z=0$ and increasing clockwise to $z=1.5$.  Crosses mark redshift intervals of $dz=0.1$.  The Sb and Sc tracks move off the plot at $z\sim 0.2$--0.3 and
re-enter at $z\sim 0.7$--1.0.
Stars represent Galactic stars from Gunn \& Stryker (1983).
The trapezoid is our selection window (equation~\ref{eq:bx}) for galaxies
with $1.9\simlt z\simlt 2.7$.
\label{fig:lbgbs}
}
\end{figure}

Since the targeted redshift range $1.9<z<2.7$ is similar to the
redshift range $2.6<z<3.4$ of the Lyman-break survey, we derived
our initial estimates of the $UG{\cal R}$ colors of galaxies
at $1.9<z<2.7$ from the observed colors of Lyman-break galaxies.
For this we used a sample of 70 Lyman-break galaxies that had spectroscopic
redshifts and measured photometry through the $U_nG{\cal R}JK_s$ bandpasses.
37 of the galaxies were taken from Shapley et al. (2001) and 33 from
Papovich et al. (2001).  
Following the approach outlined in those papers, each galaxy's 
photometry was fit with model spectral energy distributions (SEDs) that had a range
of star-formation histories and dust reddenings.  We found the best-fit SED
for every galaxy in the sample, redshifted the best-fit SEDs
to $z=1.5$, $z=2.0$, and $z=2.5$, and calculated their $U_nG{\cal R}$
colors as described in \S~\ref{sec:bcseds}.  This produced
an estimate of the $U_nG{\cal R}$ colors that each Lyman-break galaxy would have 
had if its redshift were $z\sim 2$ rather than $z\sim 3$ (figure~\ref{fig:lbgbs}).  Star-forming galaxies at $z\sim 2$ should
have $U_nG{\cal R}$ colors similar to these.

Useful color-selection criteria must not only find galaxies
at the targeted redshifts but also avoid those at other redshifts.
Galaxies at significantly higher redshifts will have extremely red
$U_n-G$ colors due to the strong Lyman break.  They are unlikely to
be confused with galaxies at $z\sim 2$.  The tracks on figure~\ref{fig:lbgbs}
show the colors of model galaxies at lower redshifts.  We considered
the templates Im, Sb, Sc from \S~\ref{sec:bcseds} and reddened each 
to $E(B-V)=0.15$ with dust that followed a Calzetti (1997) extinction curve.
At $0.3\simlt z\simlt 1.0$ galaxies have red $G-{\cal R}$ colors
due to the redshifted
Balmer/4000\AA\ breaks and are easily distinguished
from galaxies at $z\sim 2$.  At lower and higher redshifts
the potential for confusion is greater.  Young galaxies (type Im) at $z<0.3$
and $1.5<z<1.9$ have $U_nG{\cal R}$ colors sufficiently similar
to those of galaxies at $1.9<z<2.7$ that a clean separation is impossible
given $U_nG{\cal R}$ images alone.  Our selection criteria were designed
in large part to mitigate this contamination.

\begin{figure}[htb]
\centerline{\epsfxsize=9cm\epsffile{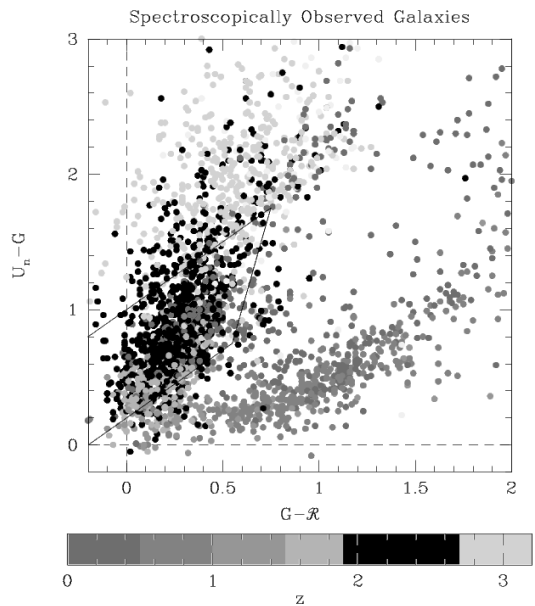}}
\figcaption[f11.eps]{
Spectroscopic redshifts of objects with different $U_nG{\cal R}$
colors.  The hooked shaped is produced by the transition
from Balmer/4000\AA\ absorption in the spectra of galaxies 
at $z\simlt 1$ to Lyman absorption in the
spectra of galaxies at $z\simgt 2.5$.  Galaxies at intermediate redshifts
have neutral $U_nG{\cal R}$ colors; those with $2.0\simlt z\simlt 2.5$
tend to lie within the region enclosed by the trapezoid (equation~\ref{eq:bx}).
\label{fig:showcandsz2}
}
\end{figure}

\begin{figure}[htb]
\centerline{\epsfxsize=9cm\epsffile{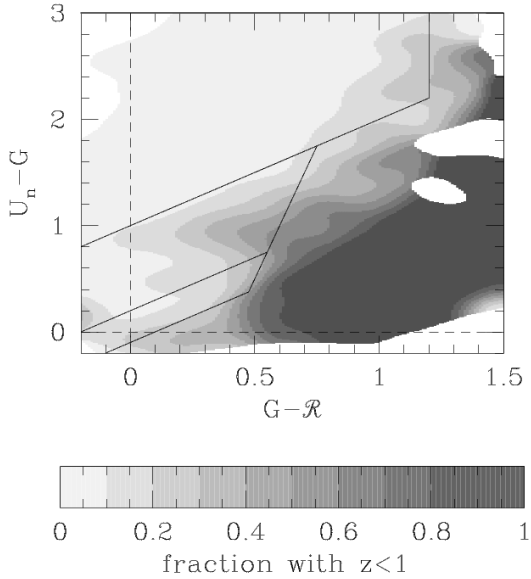}}
\figcaption[f12.eps]{ The fraction of sources that have spectroscopic redshift
$z<1$ as a function of $U_nG{\cal R}$ color.  The solid lines enclose
the Lyman-break selection window of Steidel et al. (2003) and
the two $1.4<z<2.7$ selection windows discussed below 
(equations~\ref{eq:bx} and~\ref{eq:bm}).
\label{fig:zlt1frac}
}
\end{figure}

\begin{figure}[htb]
\centerline{\epsfxsize=9cm\epsffile{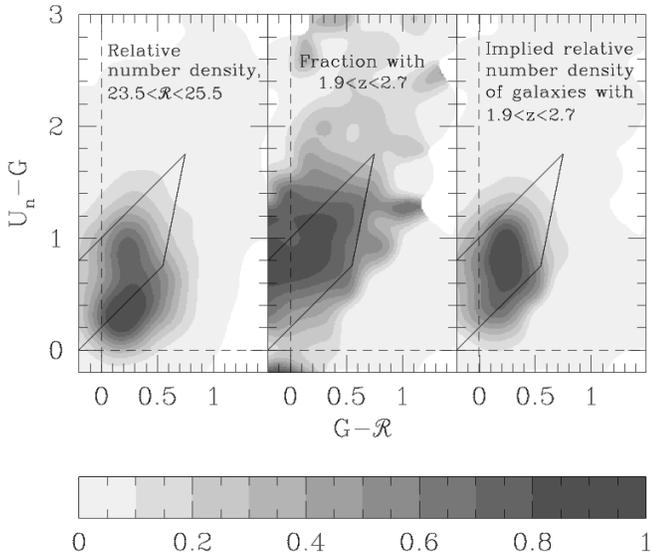}}
\figcaption[f13.eps]{Left panel:  The distribution of colors for all objects with 
${\cal R}>23.5$ in deep $U_nG{\cal R}$ images
of the HDF-N.  Middle panel:  The observed probability that an object 
with ${\cal R}>23.5$ has redshift $1.90<z<2.70$ 
as a function
of $U_nG{\cal R}$ color.  
Right panel:  The product of the left and middle panels.  This shows the
implied colors of
objects with $1.90<z<2.70$.
The trapezoid in all panels shows the selection criteria
of equation~\ref{eq:bx}.
\label{fig:z2triptych}
}
\end{figure}

To learn how to design a selection window
that would exclude as many interloping
galaxies as possible, we began in the fall of 1996
to obtain
spectra of objects with colors similar to those we expected for galaxies
at redshift $z\sim 2$.  Figure~\ref{fig:showcandsz2} shows how
galaxies' $U_nG{\cal R}$ colors depend on their redshifts.  
In addition to the galaxies
whose redshifts we measured for this purpose, the figure includes
all galaxies from the Lyman-break survey of Steidel et al. (2003) and
all galaxies from the magnitude-limited surveys 
and our Balmer-break survey
described in \S~\ref{sec:bbg}.  Figures~\ref{fig:zlt1frac} and~\ref{fig:z2triptych} 
present these
data in a way that may be easier to understand.  Both are constructed in a similar
manner to figure~\ref{fig:z1triptych}.  Figure~\ref{fig:zlt1frac}
shows the typical $U_nG{\cal R}$ colors of galaxies at $z<1$.  These
were the galaxies that
we hoped to exclude from our sample.  Figure~\ref{fig:z2triptych} shows
as a function
of $U_nG{\cal R}$ color the relative number density of sources
(left panel), the fraction of objects that lie
in the redshift range of interest $1.9<z<2.7$ (middle panel), and the implied relative number density
of sources in the same redshift range (right panel).  Because no existing
magnitude-limited surveys contain significant numbers of galaxies at
these redshifts, the right panel was calculated by
multiplying the left and center
panels together.  


The following selection criteria were inspired by the shape of the contours
on these plots, by our requirement that galaxies of all LBG-like spectral types
have a finite probability of satisfying the criteria, and by our desire to 
leave no gap between these criteria and the Lyman-break selection
criteria of Steidel et al. (2003):
\begin{eqnarray}
G-{\cal R}&\geq& -0.2\nonumber\\
U_n-G     &\geq& G-{\cal R}+0.2\quad\quad\quad\quad\quad\quad\lower 2ex\hbox{\bf [BX]}\nonumber\\
G-{\cal R}&\leq& 0.2(U_n-G)+0.4\nonumber\\
U_n-G     &<& G-{\cal R}+1.0.
\label{eq:bx}
\end{eqnarray}
These criteria are called ``BX''
in our internal naming convention and may be referred to by that name
in subsequent publications.

\begin{figure}[htb]
\centerline{\epsfxsize=9cm\epsffile{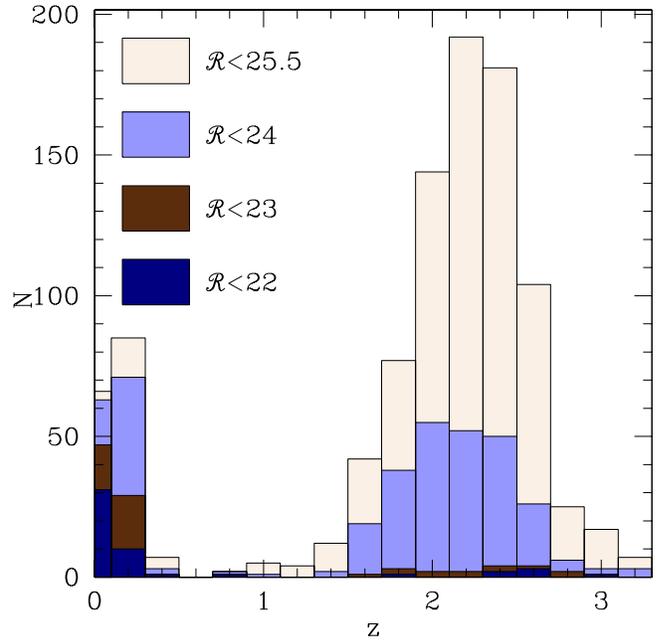}}
\figcaption[f14.eps]{
The observed redshift distribution of objects whose
colors satisfy equation~\ref{eq:bx}.
Stars are included in the bin containing $z=0$; they make up
4.6\% of the sample but only 0.9\% of the sample fainter than ${\cal R}=23.5$.
\label{fig:bxnz}
}
\end{figure}

Figure~\ref{fig:bxnz} shows the redshift distribution of randomly selected
objects whose colors satisfy these criteria.  Excluding sources
with $z<1$, the mean redshift is $\bar z= 2.22$ and the standard deviation
is $\sigma_z=0.34$.   

As could be anticipated
from figure~\ref{fig:lbgbs}, the Balmer break in low redshift galaxies 
can sometimes be confused with the Lyman-$\alpha$ forest absorption
in the spectra of galaxies at $z\sim 2$.  The resulting contamination
of our sample is severe at magnitudes ${\cal R}<23$ but negligible
by ${\cal R}=25.5$.  Restricting the sample to ${\cal R}>23.5$ provides
a crude but effective way of eliminating low-redshift interlopers.

One can roughly estimate the completeness coefficients $\alpha$ and $\beta$
for the sample as follows.  Let $P(x,y)$ be the probability
that a randomly chosen galaxy with $23.5<{\cal R}<25.5$ and
colors $G-{\cal R}=x$ and $U_n-G=y$ 
has a redshift that lies in the range $1.9<z<2.7$, let $n(x,y)$ be
proportional to the observed number density of galaxies with $23.5<{\cal R}<25.5$
that have colors $G-{\cal R}=x$ and $U_n-G=y$, and let $I(x,y)$ be
equal to 1 if the color $G-{\cal R}=x$, $U_n-G=y$ satisfies 
equation~\ref{eq:bx} and to 0 otherwise.  Then the probability than an
object with $23.5<{\cal R}<25.5$ 
at $1.9<z<2.7$ will satisfy our selection criteria is
\begin{equation}
\alpha = \int\!\! dx\,dy\,\, P(x,y) n(x,y) I(x,y)\,\,\, \Bigg/\,\,\, \int\!\! dx\,dy\,\, P(x,y) n(x,y)
\label{eq:z2alpha}
\end{equation}
and the probability that an object with $23.5<{\cal R}<25.5$ that satisfies our selection criteria
will lie at $1.9<z<2.7$ is
\begin{equation}
\beta = \int\!\! dx\,dy\,\, P(x,y) n(x,y) I(x,y)\,\,\, \Bigg/\,\,\, \int\!\! dx\,dy\,\, n(x,y) I(x,y).
\label{eq:z2beta}
\end{equation}
The functions $P(x,y)$ and $n(x,y)$ are shown in figure~\ref{fig:z2triptych}.
Numerically integrating equations~\ref{eq:z2alpha} and~\ref{eq:z2beta}
yields the estimates $\alpha=0.64$, $\beta=0.70$.  Roughly two-thirds
of galaxies with $1.9<z<2.7$ satisfy our selection criteria,
and roughly two-thirds of the galaxies that satisfy our selection
criteria lie at $1.9<z<2.7$.  The result could have been anticipated
to a large extent from the redshift histogram shown in figure~\ref{fig:bxnz}.

\section{COLOR SELECTION AT $1.4\simlt Z\simlt 2.1$}
\label{sec:bmz}
\begin{figure}[htb]
\centerline{\epsfxsize=9cm\epsffile{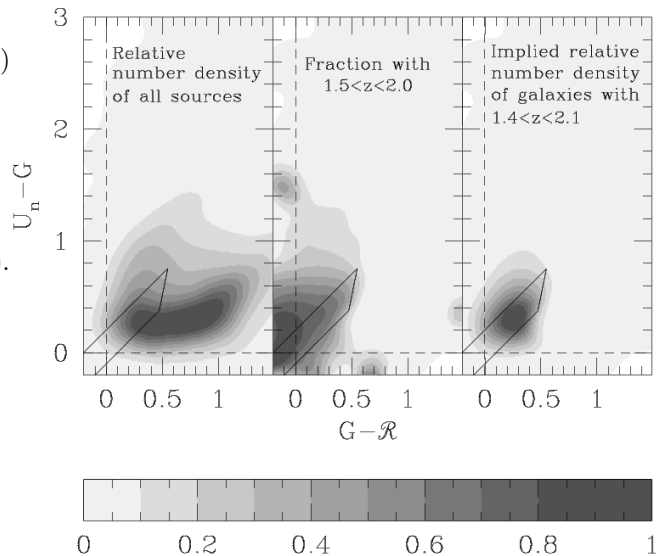}}
\figcaption[f15.eps]{
Similar to figure~\ref{fig:z1triptych}, except the targeted
redshift range is $1.4<z<2.1$, the ${\cal R}>23.5$ criteria
has been removed, and the trapezoids
show the selection criteria
of equation~\ref{eq:bm}.
\label{fig:z152triptych}
}
\end{figure}

Our approach towards defining selection criteria at these redshifts
differs little from our approach at higher redshift.  As
redshift decreases from $z=2$, galaxies become
bluer in $U_n-G$ because a smaller fraction of their
observed-frame $U_n$ emission is absorbed by the Lyman-$\alpha$ forest.
Otherwise their optical colors are largely unchanged.
One would expect galaxies with $1.4<z<2.1$
to lie just below our $1.9<z<2.7$ selection window 
(equation~\ref{eq:bx}) in the $U_nG{\cal R}$ plane.  
We began exploratory spectroscopy of galaxies in this
part of the $U_nG{\cal R}$ plane in the fall of 1997.
Observations continued sporadically until the spring of 2003.
Figure~\ref{fig:z152triptych} shows that these observations
largely conformed to our expectations.  
The following selection criteria were inspired by the shape of the contours
on this plot, by our requirement that galaxies of all LBG-like spectral types
have a finite probability of satisfying the criteria (see
figure~\ref{fig:lbgbs}), and by our desire to 
leave no gap between these criteria and the selection
criteria of equation~\ref{eq:bx}:
\begin{eqnarray}
G-{\cal R}&\geq& -0.2\nonumber\\
U_n-G     &\geq& G-{\cal R}-0.1\quad\quad\quad\quad\quad\quad\lower 2ex\hbox{\bf [BM]}\nonumber\\
G-{\cal R}&\leq& 0.2(U_n-G)+0.4\nonumber\\
U_n-G     &<& G-{\cal R}+0.2.
\label{eq:bm}
\end{eqnarray}
These criteria are called ``BM''
in our internal naming convention and may be referred to by that name
in subsequent publications.

\begin{figure}[htb]
\centerline{\epsfxsize=9cm\epsffile{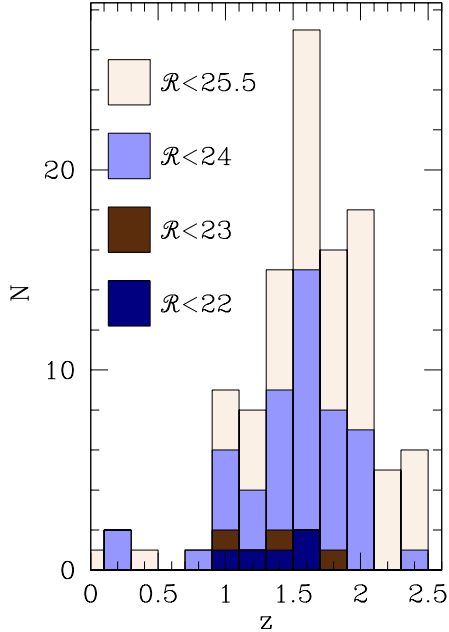}}
\figcaption[f16.eps]{
The redshift distribution of randomly selected objects whose
colors satisfy equation~\ref{eq:bm}.
\label{fig:bmnz}
}
\end{figure}

\begin{figure}[htb]
\centerline{\epsfxsize=9cm\epsffile{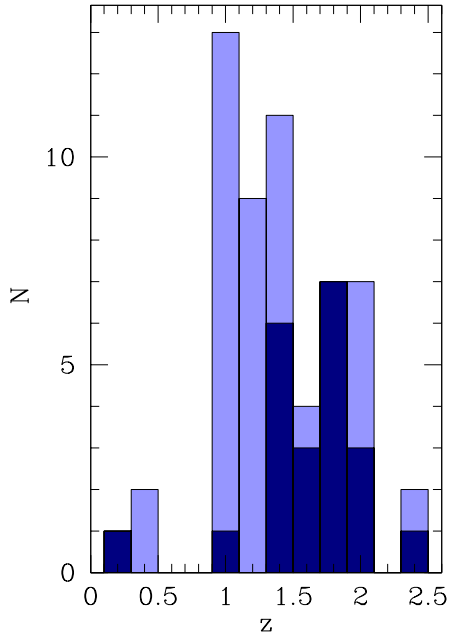}}
\figcaption[f17.eps]{
Light histogram: the redshift distribution of spectroscopically observed objects in
the HDF-N whose
colors satisfy equation~\ref{eq:bm}.  The stellar contamination rate
is about 1\%.
Dark histogram: the redshift distribution of HDF-N objects
whose colors satisfy equation~\ref{eq:bm} but
not equation~\ref{eq:fnz15}.
\label{fig:bmnzhdf}
}
\end{figure}

The completeness coefficients $\alpha$ and $\beta$ for these selection
criteria, estimated with the approach of equations~\ref{eq:z2alpha}
and~\ref{eq:z2beta}, are listed in table~\ref{tab:z2compint}.
Roughly $42$\% of galaxies with $1.4<z<2.1$ and ${\cal R}<25.5$ satisfy the
selection criteria, and roughly $46$\% of the objects with ${\cal R}<25.5$
that satisfy the criteria are galaxies with $1.4<z<2.1$.
Figure~\ref{fig:bmnz} shows the observed  redshift distribution of the sources
whose colors satisfied equation~\ref{eq:bm}.   Excluding sources with $z<1$,
the mean redshift is $\bar z=1.70$ and the standard deviation
is $\sigma_z=0.34$.  As figure~\ref{fig:lbgbs}
shows, galaxies' $U_nG{\cal R}$ colors do not change by a large
amount between $z=1.0$ and $z=1.5$, and as a result the redshift
distribution has a significant tail extending to $z\sim 1$.  This
low-redshift tail can be eliminated to a large extent, if $z$ band
photometry is available, by excluding from the sample
any objects whose colors satisfy the $1.0<z<1.5$ selection criteria
of equation~\ref{eq:fnz15}.  See figure~\ref{fig:bmnzhdf}.  Subsequent papers
may refer to this combination of color-selection criteria as ``BMZ.''

\begin{deluxetable}{lc
}\tablewidth{0pc}
\scriptsize
\tablecaption{Color-selection Efficiency, $1.4\simlt z\simlt 2.7$}
\tablehead{
	\colhead{Redshift} &
        \colhead{$\alpha,\beta$\tablenotemark{a}} 
}
\startdata
$1.4<z<2.1$:& 0.42,0.46 \\
$1.9<z<2.7$:& 0.64,0.70 \\
\enddata
\tablenotetext{a}{The estimated fraction of galaxies at the redshift of interest whose colors satisfy our proposed selection criteria ($\alpha$) and the estimated fraction of objects satisfying our selection criteria whose redshift lies in the desired range ($\beta$).}
\label{tab:z2compint}
\end{deluxetable}

\section{$U_nG{\cal R}$ COLOR SELECTION AT ANY REDSHIFT $1<z<3$}
\label{sec:anyz}
\begin{figure}[htb]
\centerline{\epsfxsize=9cm\epsffile{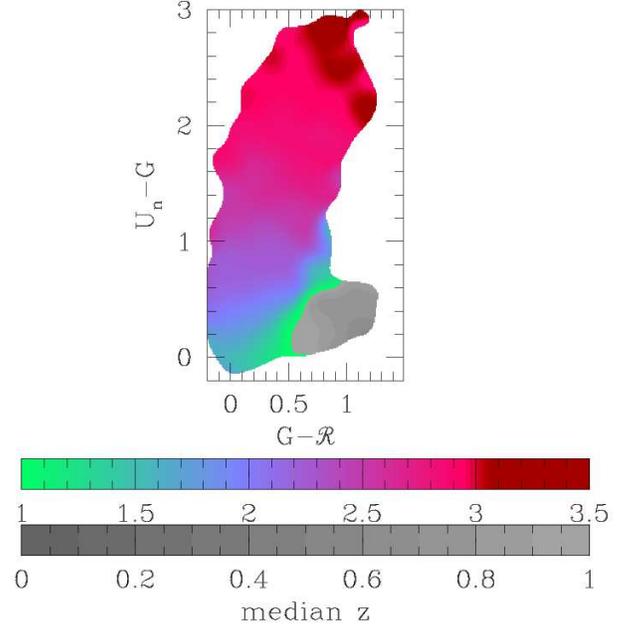}}
\figcaption[f18.eps]{Median redshift of spectroscopic sources as a function of
$U_nG{\cal R}$ color.  Galaxies from the $U_nG{\cal R}$-selected
spectroscopic sample described
in this paper, from the Lyman-break survey of Steidel et al. (2003),
and from the magnitude-limited surveys listed in \S~\ref{sec:bbg}
were placed into bins with $\Delta(U_n-G)=0.1$,
$\Delta(G-{\cal R})=0.1$ and the median redshift was calculated
for each bin.  The resulting array of median redshift vs.
$U_nG{\cal R}$ color was then smoothed by a Gaussian with
$\sigma_{U_n-G}=\sigma_{G-{\cal R}}=0.1$ to produce this plot.
The slow change of median redshift with color makes it easy
to devise selection criteria that target arbitrary redshift
ranges within the larger span $1<z<3$.
\label{fig:zmedian}
}
\end{figure}

The previous two sections presented color-selection criteria tuned
to the two arbitrary redshift ranges $1.4<z<2.1$ and $1.9<z<2.7$.
Some projects may require a sample of galaxies with a similar
but slightly different range of redshifts, e.g., $1.7<z<2.3$.  
Minor adjustments to the selection criteria we have presented  
can tune them to this redshift range or others.  To help readers
estimate how to adjust our criteria to produce samples
with a desired range of redshifts, we show in figure~\ref{fig:zmedian}
the observed median redshift as a function of $U_nG{\cal R}$ color
of galaxies with ${\cal R}>23.5$ 
in our spectroscopic sample.  
Roughly aligning the edges of a selection box with
this plot's contours 
will produce reasonably good selection criteria tuned
to arbitrary redshifts within the interval $1\simlt z\simlt 3$.
Some spectroscopic follow-up will be required to
verify the median redshift of the sources and to place
limits on the sample's contamination by stars and low-redshift galaxies.

\section{SUMMARY AND DISCUSSION}
\label{sec:summary}

It is often asserted that studying galaxies at
redshifts $1<z<3$ will be tremendously difficult.
The optical spectra of galaxies near the middle of this redshift range 
contain neither the strong spectral breaks that are sometimes thought
to be necessary for effective photometric selection
nor the strong emission lines Ly-$\alpha$ or [OII]$\lambda 3727$ 
that are sometimes thought to be crucial for spectroscopic identification.
The supposed difficulty of galaxy observations causes
many to refer to this redshift range as the spectroscopic desert
or high place of sacrifice (Bullock et al. 2001).

\begin{figure}[htb]
\centerline{\epsfxsize=9cm\epsffile{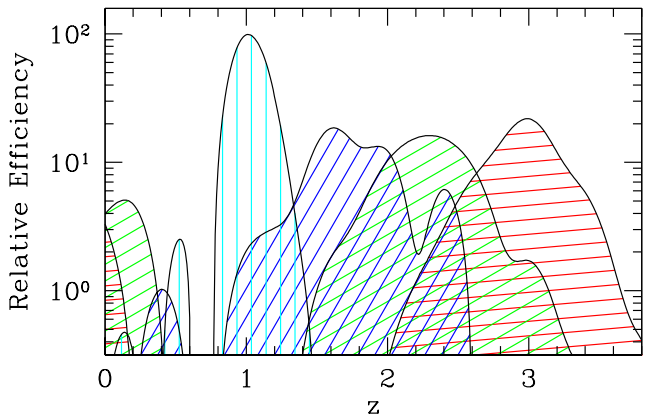}}
\figcaption[f19.eps]{Scaled redshift distributions of objects selected
with equations~\ref{eq:fn}, \ref{eq:bm}, and~\ref{eq:bx}, 
and with the $z\sim 3$ Lyman-break selection criteria of
Steidel et al. (2003).  The actual redshift histogram of
each sample was divided by the spectroscopic observing
time devoted to the sample.  All histograms were then multiplied
by the same arbitrary constant and fit with a cubic spline.
The area under each curve is proportional to the number of
spectroscopic redshifts we measured per hour with
the LRIS spectrograph.  Galaxy
spectroscopy is not much harder at $1<z<3$ than at
higher or lower redshifts---as long as the spectrograph has good UV 
throughput.
\label{fig:zperhour}
}
\end{figure}

We are sceptical.  The optical colors of galaxies at redshifts $1<z<3$ 
{\it are} distinctive.  We showed in sections~\ref{sec:bbg},~\ref{sec:bx},
and~\ref{sec:bmz} that these galaxies
are easy to locate in deep images using 
simple color-selection criteria
(equation~\ref{eq:fn} for $0.9<z<1.1$, equation~\ref{eq:fnz15} for $1.0<z<1.5$, 
equation~\ref{eq:bx} for $1.9<z<2.7$, and equation~\ref{eq:bm} for $1.4<z<2.1$).
Once they have been found their redshifts are no harder to measure
than those of comparably bright galaxies at slightly higher or lower
redshift.  This is due in large part to the strength of
their interstellar absorption lines.
With an appropriately chosen spectroscopic set-up,
redshifts are as easy to measure from absorption lines
as from the Lyman-$\alpha$ or [OII]$\lambda$3727 emission lines.
The point is illustrated by figure~\ref{fig:zperhour}; during our
surveys we obtained as many redshifts per hour of observing
time at $1.5<z<2.5$ as at the higher redshifts where Lyman-$\alpha$
is more easily observed.  That was possible only because we knew
the approximate redshifts of our sources in advance and could
optimize our choice of spectrograph and its configuration accordingly.
We could not have measured a redshift for many of the galaxies at $1.5<z<2.5$ 
with a spectrograph that lacked the good UV throughput
of LRIS-B.
See Steidel et al. (2004) for a more detailed
discussion.

\begin{figure}[htb]
\centerline{\epsfxsize=9cm\epsffile{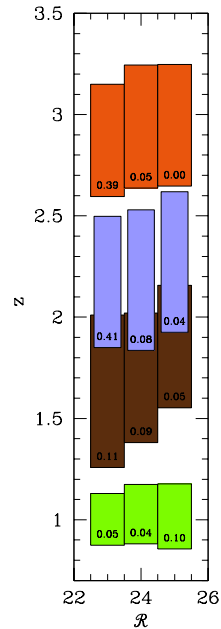}}
\figcaption[f20.eps]{Redshift distributions as a function of apparent
magnitude.  Shaded regions show the mean $\pm 1\sigma$ redshift range for
objects of different apparent magnitudes whose colors satisfy
(from bottom to top) equations~\ref{eq:fn}, \ref{eq:bm}, \ref{eq:bx}
and the $z\sim 3$ Lyman-break selection criteria of Steidel et al. (2003).
The $1\sigma$ interval was calculated after excluding interlopers
with $z<0.4$ from the $z\simeq 1.0$ sample and interlopers with $z<1$
from the other samples.  The number in each box indicates the fraction
of sources that were excluded.  This fraction
depends strongly on magnitude for some samples, but otherwise
the redshift distributions are insensitive apparent magnitude.
It is therefore reasonable to assume, when the interloper fraction is small,
that faint photometric candidates will have the same redshift distribution
as brighter candidates with spectroscopic redshifts.
The weak trend of higher redshifts for (apparently) fainter objects results 
from the change in luminosity distance.
\label{fig:meanz_vs_r}
}
\end{figure}

Two to three nights spent imaging a single field with a UV-sensitive
$40'\times 40'$ camera on a 4m telescope is sufficient
to detect $\sim 2\times 10^4$ galaxies that satisfy
one set of the $1<z<3$ color-selection criteria that we have presented.
A major benefit of color-selected spectroscopy
is that it lets one draw statistically significant conclusions
from this large number of high-redshift galaxies without having
to measure a redshift for every one.  For example, the $\sim 1000$ 
spectroscopic redshifts
we have measured for objects whose colors satisfy
equation~\ref{eq:fn} tell us, with high precision, what the redshift
distribution must be for the large ensemble of objects in
the $40'\times 40'$ image whose colors satisfy equation~\ref{eq:fn}.
One can use this knowledge to derive the luminosity function
or spatial clustering strength of galaxies at $z\sim 1$
from the list of photometric candidates alone.
Because the redshift distribution of photometric candidates
does not depend strongly on magnitude for faint galaxies
(figure~\ref{fig:meanz_vs_r}), 
it should be possible to use purely photometric observations
to learn about the properties of the numerous high-redshift
galaxies that are too faint for spectroscopy.  
This may be best use of our $z\sim 1$ color-selection criteria,
since the brightest galaxies at this redshift are routinely detected
in the large and ongoing spectroscopic surveys of
Davis et al. (2003) and Le Fevre et al. (2003).

The weakness of color-selected surveys is that not all galaxies
at the targeted redshifts will satisfy the adopted selection
criteria.  
A high level of completeness at the targeted redshifts can be
obtained only at the price of admitting large numbers of galaxies
at the wrong redshifts.
With the simple selection criteria we presented, observers
will find more than $\sim$ one-half of the galaxies brighter than the magnitude
limit at the redshift of interest
and will waste no more than $\sim$ one-half of their observing
time observing objects at other redshifts.  50\% completeness
is not ideal, but the problem is less severe than one might
imagine.   First, a star-forming galaxy at $1<z<3$ that does not
satisfy one of our color-selection criteria will likely satisfy
another.  Consider galaxies with redshift $1.8<z<2.2$
and magnitude ${\cal R}<25.5$, for example.
Equation~\ref{eq:z2alpha} implies that only 63\% of them will
satisfy the $1.9<z<2.7$ selection criteria of equation~\ref{eq:bx}.
But an additional 26\% will satisfy the $1.5<z<2.0$ selection criteria
of equation~\ref{eq:bm}, and an additional 5\% will satisfy the
Lyman-break galaxy selection criteria of Steidel et al. (2003).
A total of 94\% of photometrically detected
galaxies with $23.5<{\cal R}<25.5$ and $1.8<z<2.2$ will satisfy
one of the $U_nG{\cal R}$ selection criteria we have presented.
By conducting redshift surveys with each of these criteria,
one can reduce the incompleteness to a reasonably low level.
Second, even if a survey adopts only one of our selection criteria,
much of the incompleteness can be corrected in a statistical sense.
The incompleteness is
largely due to the photometric errors in our color measurements, which
are not small compared to the size of our color-selection
window.  Many galaxies whose true colors lie inside our selection
window will have measured colors that lie outside the window;
many galaxies whose measured colors lie inside 
will have true colors that lie outside. 
The result
is a broad, bell-shaped redshift histogram rather than
a boxcar extending from the minimum to the maximum targeted
redshift.  Because photometric uncertainties are easy
to characterize with Monte-Carlo simulations, their contribution
to our incompleteness is easy to understand and correct.
Adelberger (2002,2004) explains in detail.

In any case, all observational strategies require some compromise
between efficiency and completeness, and the compromises of 
color selection do not look bad compared to the alternatives.  
Suppose one were interested in studying galaxies at
$0.85<z<1.15$.  By obtaining a spectrum of every galaxy
brighter than ${\cal R}=24.0$ in an image one would be
certain to produce a statistically complete sample of
galaxies at $z\sim 1$ to this magnitude limit, but over 80\%
of the observing time would have been spent
on objects at the
wrong redshifts (e.g., Cohen et al. 2000).
If one were willing to tolerate 15\% incompleteness in observed $U_n$
luminosity density, one could cull spectroscopic targets with
the color criteria of equation~\ref{eq:fn} and reduce the
required observing time by more than a factor of three 
(see table~\ref{tab:fncompint}).  As redshift increases
the benefits of color-selection become more obvious.
Only one object out of 50 in the magnitude limited survey
of Cohen et al. (2000) had a redshift $1.9<z<2.7$.  If one
wished to study galaxies at these redshifts with a magnitude
limited survey the required observing time would be
33 times longer than if one color-selected targets with
equation~\ref{eq:bx} (see table~\ref{tab:z2compint}).  
A project that could be completed in 1--2 years
with color-selection would require a lifetime of observing
with standard magnitude-limited techniques. The $\sim 30$\% incompleteness of
color-selected surveys does not seem a high price to pay.\footnote{A
magnitude-limited survey conducted with a blue-optimized spectrograph
would find a higher fraction of sources with $1.9<z<2.7$ than
Cohen's 1 in 50, so the comparison is somewhat unfair.  In practice, however,
few would chose to undertake a magnitude-limited survey with a blue
spectrograph because it would make spectra more difficult to identify
at the lower redshifts where most sources lie.  The ability to take full
advantage of optimized spectrographs is a non-negligible benefit
of color selection.}

Magnitude-limited optical surveys are not the only alternative
to color-selected optical surveys.  Many have
advocated finding galaxies at $1<z<3$ with photometry
outside of the optical window.  We chose to develop criteria
that relied solely on optical photometry because ground-based
optical imagers offer an unrivaled combination of high sensitivity,
high spatial resolution, and large fields-of-view.  This is
an advantage that is difficult to overcome.  
Surveys at other wavelengths have their strengths---near-IR observations
should provide a more complete census of older stars (e.g., Franx et al. 2003,
Rudnick et al. 2003) and far-IR/sub-mm observations should yield more
reliable estimates of star-formation rates---but the fact remains
that an investigator given two nights on an older 4m optical telescope
can create a photometric sample of high-redshift galaxies 
larger than all existing samples at other wavelengths combined.
It seems likely to us that much of
what we will learn about galaxies at $1<z<5$ in the coming decade
will come from large optical surveys selected with color criteria similar
to ours.

Attentive readers may have recognized that the derivation of these
criteria
did not require much thought.  That is exactly the point.
The value of this paper, if any, lies not in our photometric selection
criteria themselves but in the proof
that large samples
of galaxies at $1<z<3$ can be created with trivial techniques that
rely solely on ground-based imaging and well-developed CCD technology.
The spectroscopic desert is a myth.

\medskip

KLA is deeply grateful for unconditional support from
the Harvard Society of Fellows.  It will be missed. 
CCS and AES were supported by grant AST0070773
from the U.S. National Science Foundation and by
the David and Lucile Packard Foundation.
NAR is supported by the National Science Foundation.
This research made use of the NASA/IPAC Extragalactic Database (NED),
which is operated by the Jet Propulsion Laboratory, California Institute
of Technology, under contract with the National Aeronautics and Space
Administration.
The authors wish to extend special thanks to those
of Hawaiian ancestry for allowing
telescopes and astronomers upon their sacred mountaintop.
Their hospitality made our observations possible.

\bigskip

\end{document}